# Formation of Episodic Magnetically Driven Radiatively Cooled Plasma Jets in the Laboratory

F. Suzuki-Vidal[1], S. V. Lebedev[1], A. Ciardi[2], S. N. Bland[1], J. P. Chittenden[1], G. N. Hall[1], A. Harvey-Thompson[1], A. Marocchino[1], C. Ning[1], C. Stehle[2], A. Frank[3], E. G. Blackman[3], S. C. Bott[4], T. Ray[5]

[1] The Blackett Laboratory, Imperial College, London, SW7 2BW, UK.
[2] LERMA, Observatoire de Paris, CNRS and UPMC, Meudon 92195 France.
[3] Department of Physics and Astronomy, University of Rochester, Rochester, NY, USA.
[4] Center for Energy Research, University of California, San Diego, 92093-0417, USA.
[5] Dublin Institute for Advanced Studies, Dublin, Ireland.

**Abstract.** We report on experiments in which magnetically driven radiatively cooled plasma jets were produced by a 1 MA, 250 ns current pulse on the MAGPIE pulsed power facility. The jets were driven by the pressure of a toroidal magnetic field in a "magnetic tower" jet configuration. This scenario is characterized by the formation of a magnetically collimated plasma jet on the axis of a magnetic "bubble", confined by the ambient medium. The use of a radial metallic foil instead of the radial wire arrays employed in our previous work allows for the generation of episodic magnetic tower outflows which emerge periodically on timescales of ~30 ns. The subsequent magnetic bubbles propagate with velocities reaching ~300 km/s and interact with previous eruptions leading to the formation of shocks.

**Introduction.**

Highly collimated jets and outflows from protostars have been of increasing interest to observational and theoretical astrophysics as observational techniques and computer simulations continue to improve. Simultaneously a new branch of high-energy density physics – laboratory astrophysics – has been able to reproduce the dynamics shown on the stellar scale within the laboratory, maintaining the relevant dimensionless parameters (i.e. Mach number, Reynolds number, cooling parameter, etc.) given by the MHD scaling laws (Ryutov et al. 2000). Experimental facilities have shown to be capable of performing experiments which reproduce particular features of these objects by using high currents in a z-pinch machine (Lebedev et al., 2002; Lebedev et al., 2005b) and high power lasers (Farley et al., 1999; Foster et al., 2002; Blue et al., 2005).

Different models of the formation of protostellar jets have proposed that magnetic fields are responsible for driving and collimating outflows from a system composed by a star with an accretion disk (Blandford & Payne, 1982). In a particular scenario the magnetic field topology evolves, due to differential rotation, into one with a predominantly toroidal magnetic field, which collimates ejected material from the system as a jet on the axis of a cavity confined by the external ambient pressure. This "magnetic tower" model (Lynden-Bell, 1996) has been proposed as a mechanism of jet formation for different astrophysical objects ranging from protostars to neutron stars (Lynden-Bell, 2006; Goodson et al., 1999; Uzdensky and MacFayden, 2006; Kato et al., 2004). The experimental approach that reproduces some aspects of the plasma jet dynamics relevant to this model has been the radial wire array z-pinch configuration (Lebedev et al., 2005a), in which a plasma jet is collimated and driven onto the axis of a magnetic "bubble" by rising toroidal magnetic field loops. Magneto-hydrodynamic simulations have shown that dimensionless parameters in these experiments are relevant to jets from young stellar objects (Ciardi et al., 2007).

In this paper we present experiments in which episodic formation of magnetic tower jets was observed. The experimental set-up also allows us to vary controllably the density of the ambient medium through which the magnetic tower jets propagate. It is believed that the knots and shocks observed in the protostellar jets could originate from both the variability of the outflow at the jet formation stage, or could arise from the interaction with the ambient medium. The experimental capabilities developed in the present work can contribute to a better understanding of the issues related to variability of astrophysical jets.

**Experimental setup.**

The experimental configuration is similar to the radial wire array z-pinch used in our previous experiments (Lebedev et al., 2005a). In the present experiments the current from the MAGPIE generator (peak current of 1MA in 250ns) (Mitchell at al., 1996) is driven into a 6 - 6.5 μm thick aluminum foil, which is held radially between two concentric electrodes (Fig. 1). The central electrode (cathode) is a hollow cylinder with a diameter of 3.1 mm, with the diameter of the outer electrode being 60 mm. Diagnostics included: laser probing ($\lambda$=532 nm, $t$~0.4 ns) providing 2-frame interferometry, shadowgraphy and schlieren imaging; time resolved (~2 ns exposure) pinhole cameras which recorded emission in the XUV region (>30 eV) providing up to 8 frames per experiment; magnetic "pick-up" probes to measure any trapped magnetic field inside the outflows; an inductive probe connected to the cathode to measure voltage and thus Poynting flux driving the outflow.

The imposed current path (Fig. 1) produces a toroidal magnetic field $B_\phi$ below the foil which is directly proportional to the current and decreases with the radial distance from the cathode ($B_\phi \propto I(t)/r$). For peak current the toroidal magnetic field can reach magnitudes of $B_\phi \sim 100$ T (1 MG) at the cathode radius.

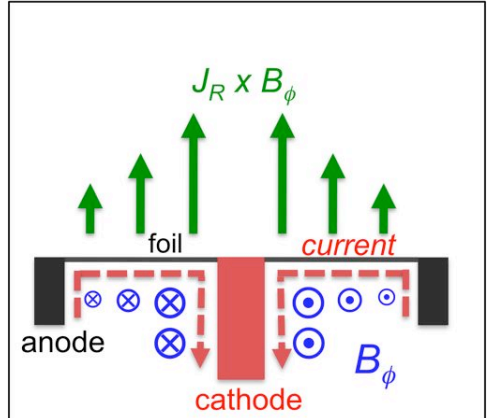

Fig. 1: Schematic of the radial foil setup showing the current path through the foil ($J_R$), toroidal magnetic field ($B_\phi$) and the net $J_R \times B_\phi$ force acting on the plasma produced from the foil.

As the current increases in time the foil is ohmically heated by the current, showing evidence that is consistent with the foil being melted by the current together with the formation of plasma on the foil surface. The ablated plasma flows from the foil surface in the axial ($J_R \times B_\phi$) direction. Side-on laser probing imaging taken at 172 ns from the current start (Fig. 2), shows an axial displacement of the foil near the central electrode, which is reasonably well described by 0-D equations of motion, assuming that most of the foil mass is accelerated by the pressure of the toroidal magnetic field. The material ablated from the foil fills the region above it with a low-density background plasma with a typical electron density integrated along the laser line of sight measured by laser interferometry of $N_e \sim 10^{18}$ cm$^{-2}$. Axial density profiles reconstructed from interferometry show an exponential decay with the height above the foil, suggesting an isothermal expansion of the plasma with a typical sound speed of $C_s \sim 9$ km/s.

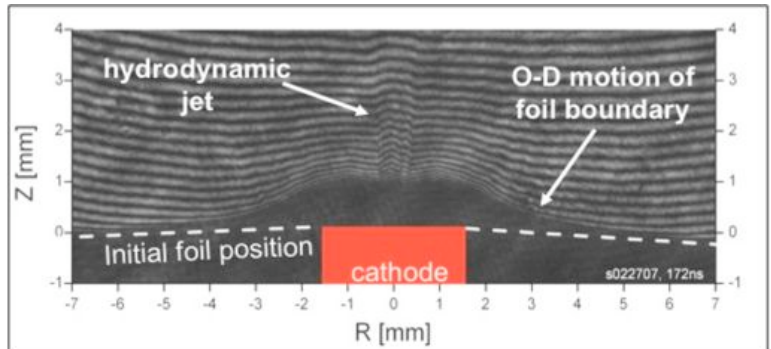

Fig. 2: Side-on laser interferometry of the radial foil at 172 ns. The pressure of the toroidal magnetic field below the foil starts displacing plasma axially and a hydrodynamic jet is formed on axis from plasma ablated from the foil.

Fig. 2 also shows an enhancement of density in the ablated plasma in the region near the axis. This hydrodynamic "precursor" jet is formed from plasma ablated from the surface of the foil, which is redirected towards the axis by radial pressure gradients. The formation of such jets by converging plasma flows has been also reported in previous laboratory astrophysics experiments using conical wire arrays and radial wire arrays (Ampleford et al., 2008; Lebedev et al., 2002; Lebedev at al., 2005b). When an ambient gas was injected into the region above the foil in the present experiments, the presence of this hydrodynamic jet affected the early time dynamics of the interaction of the outflow with the ambient, as will be discussed later in this paper.

**Formation of episodic magnetic tower jets**

The formation of magnetically driven jets starts later in time, when the Lorentz $J_R \times B_\phi$ force (which is strongest at the cathode radius) leads to ablation of all of the foil mass near the cathode and to the formation of a small radial gap between the cathode and the remainder of the foil. From this moment the Poynting flux can be injected through this gap into the region above the foil. The toroidal magnetic field pushes the ablated plasma axially and radially outwards and also pinches the plasma on axis, forming a magnetic tower jet configuration. At this stage the current flows along the jet on the axis of the magnetic cavity and along the walls of the cavity, in the same way as in our previous experiments (Lebedev et al., 2005a, Ciardi et al., 2007). The magnetic pressure from these rising toroidal loops inside the cavity inflates it both radially and axially, with measured velocities of $V_R$~50-60 km/s and $V_Z$~130-200 km/s respectively. Experimental results showing such dynamics are shown in Fig. 3. It can be seen that the initial diameter of the bubble is given by the diameter of the cathode. The most prominent feature of this new experimental set-up is that we now observe several subsequent outflows formed in the same experiment. It is possible to follow the axial positions of the subsequent episodes of the outflows shown in Fig. 3, with Fig. 4 presenting the measurements that allowed the determination of their velocities along the axis. It is seen that each outflow is expanding with approximately constant velocity, and the extrapolation of the trajectories back in time allows determining the starting time for each episode. Each subsequent bubble expands with a faster velocity, reaching $V_Z$=325 km/s for the third observed magnetic cavity. This increase in velocity is consistent with sweeping of the ambient plasma by the earlier episodes, thus allowing the subsequent magnetic bubbles to propagate through a lower ambient density. Fig. 4 also shows that the episodic outflows are accompanied by episodic outbursts of soft x-rays (photon energy between 200-300 eV and above 800 eV), which can be well correlated with the formation of each new magnetic tower jet. This is an indication that each new episode starts from the pinching of plasma on the axis of the magnetic cavity and that pinched plasma is the source of the x-ray emission. Both the axial expansion dynamics and the periodicity of x-ray emission show a timescale of ~30 ns for the formation of subsequent magnetic tower outflows.

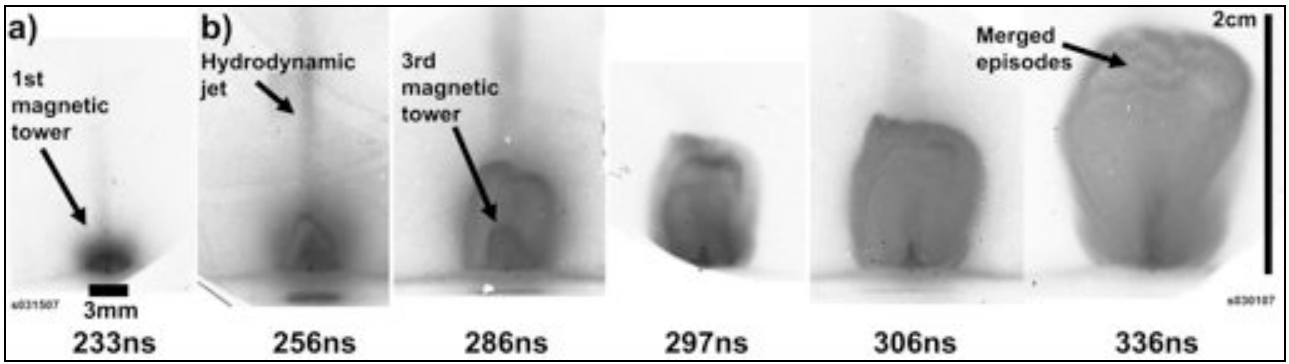

Fig. 3: Side-on XUV emission images showing the formation of episodic magnetic towers outflows which emerge with a periodicity of ~30 ns. The sequence shows results from a single experiment with the exception of the image at 233 ns.

The formation of episodic magnetic tower outflows occurs due to reconnection of current at the base of the cavity, as the gap formed between the foil and the central electrode is closed by plasma. The mass distribution in the radial foil set-up is different from that in radial wire arrays used in our previous experiments (~r for a foil and constant when using wires). This could lead to a smaller gap formed in the foil at the start of the first outflow and for a faster closure of the gap by the plasma expanding from the cathode and the remaining foil. The typical width of this gap estimated from the shape of the magnetic cavity walls is $\Delta r$~0.3-0.7 mm. The $J_R \times B_\phi$ force acting on the plasma closing the gap will push this plasma upwards and will lead to formation of a new magnetic tower outflow. The process of gap closure and formation of new outflows will continue for the duration of the current pulse from the generator, allowing to obtain 4-5 episodes for each single experiment. We should note that preliminary results show that episodic jets can also be obtained by using radial wire arrays with a high wire number.

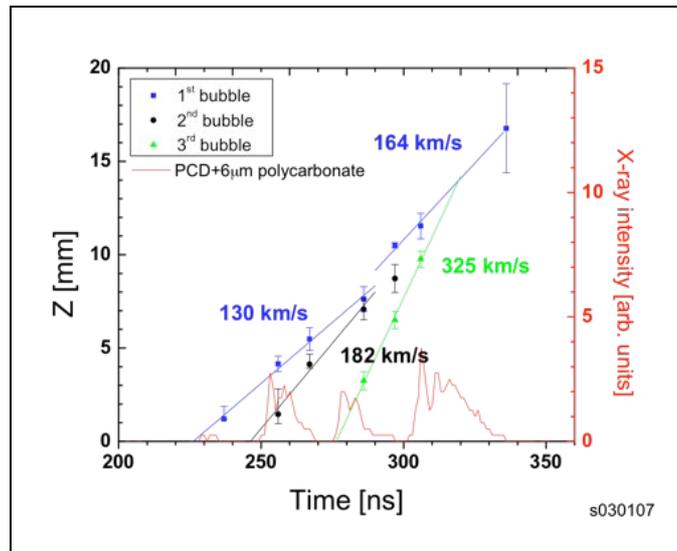

*Fig.4: Measurements of the axial extent of the episodic magnetic bubbles for the sequence in Fig.3. Each episode is correlated with soft x-ray emission produced by the pinching of a plasma jet on axis.*

**Jet propagation in an ambient gas**

The radial foil setup readily allows for the addition of an ambient neutral gas, which is injected via a supersonic gas nozzle into the space above the foil before the start of the current pulse, in a similar way to previous experiments with conical wire arrays (Ampleford, 2005). Argon was used in most of the experiments although other gases (e.g. Xe, He) were also tested. In the experiments reported here an estimated initial number density of the argon gas was $N \sim 10^{17}$-$10^{18}$ cm$^{-3}$.

The presence of the ambient gas above the foil led to several new features in the jet formation and propagation. At early times the interaction of the hydrodynamic jet and the plasma ablated from the foil with the background gas led to the formation of a conical shock with an opening angle (measured from the jet axis to the conical shock boundary) of ~60° moving with an axial velocity of $V_Z \sim 60$ km/s (Fig. 5a). It was observed that later in time (at ~260 ns) this conical shock splits into two oblique shock structures. The reasons for this phenomenon are still under investigation. It is also observed that ahead of the tip of the hydrodynamical jet a second shock feature is formed at ~230 ns. This bow shock, best seen in XUV images, is a spherical front moving at a faster velocity of $V_Z \sim 110$ km/s (Fig. 5a-5b). The interaction of this bow shock with the initial conical shock forms a contact boundary ("Mach stem") which can be seen in Fig. 5b as a horizontal dark (emitting) line. It is possible that radiation from the working surface at the end of the hydrodynamic precursor jet is playing a role in the formation of this fast bow-shock like structure, but a more detailed investigation is needed here. The addition of an ambient gas shows no significant effect on the periodicity of magnetic tower jet formation. Fig. 5c-5d show the evolution of an embedded magnetic tower jet formed inside the remnants from the earlier outflow episodes. For this embedded magnetic tower both the envelope and the central jet are clearly seen. As the subsequent magnetic bubbles reach axial expansion velocities of ~300 km/s they can catch-up and interact with the initial shock features produced by the earlier episodes. This interaction could be responsible for the complex shock structure seen at the top of the cavity in Fig. 5d.

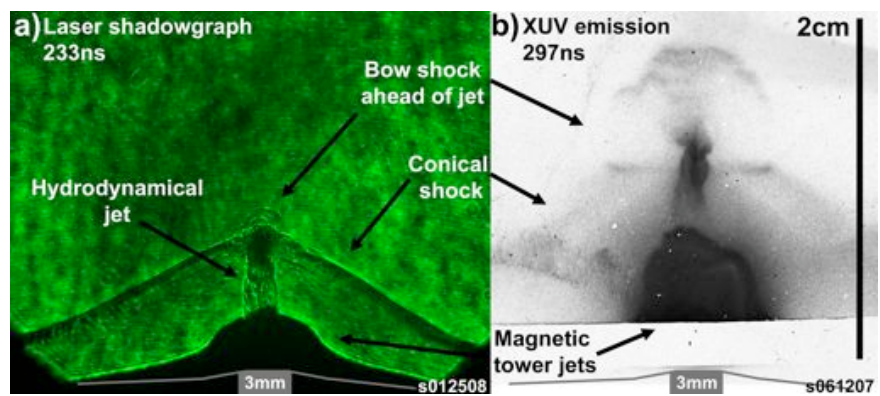

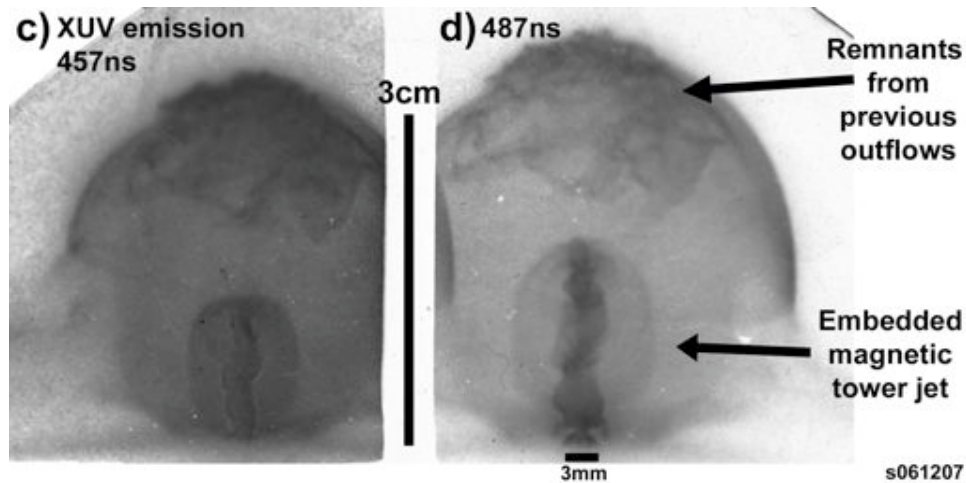

*Fig. 5: Images showing the dynamics of the interaction of jets from a radial foil with an argon gas background. An initial conical shock is formed by the hydrodynamical jet and the plasma ablated from the foil. A faster propagating bow-shock-like structure develops ahead of the hydrodynamical jet (Figs. a-b). Magnetically driven jets emerging later in time overtake these earlier formed structures (Figs. c-d).*

**Future directions**

The results of radial foil experiments show for the first time a way of producing episodic magnetically driven jets in the laboratory. The similarities of the dynamics of these episodic jets with the single-episode magnetic tower jets studied in our previous experiments with radial wire arrays (Lebedev et al., 2005a, Ciardi et al., 2007) indicate that, although some of the plasma parameters in the radial foil experiments are yet to be accurately measured, the dimensionless numbers are expected to be similar and therefore these new experiments are relevant to the physics of young stellar objects-jet launching. Preliminary measurements of trapped toroidal magnetic field inside the magnetic towers ($B_\phi \sim 0.5$ T) indicate a sufficiently high magnetic Reynolds number, which is consistent with a temperature of the magnetically driven jet of ~200 eV, estimated using x-ray spectroscopy. Voltage measured with an inductive probe across the gap of the magnetic towers have given first estimates of Poynting flux and magnetic energy available to drive these outflows. Detailed analysis of these results will be subject of future publications. The process of current reconnection responsible for the formation of the episodic jets in the experiments can be compared to the phenomenon of mass accretion and magnetic reconnection that could be responsible for episodic outflows in young stellar objects (Goodson et al., 1999).

**Acknowledgments**

This research was supported by the European Community Marie Curie JETSET network (contract MRTN-CT-2004 005592) and the SSAA program of the NNSA (DOE Cooperative Agreement DE-FC03-02NA00057).